\documentclass[preprints,accept,pdftex,moreauthors]{Definitions/mdpi} 

\firstpage{1} 
\pubvolume{1}
\issuenum{1}
\articlenumber{0}
\pubyear{2023}
\copyrightyear{2023}
\datereceived{ } 
\daterevised{ } 
\dateaccepted{ } 
\datepublished{ } 
\hreflink{https://doi.org/} 
\pdfoutput=1 

\Title{Physics-based signal analysis of genome sequences: \\ GenomeBits overview}

\Author{E. Canessa}

\address[1]{%
$^{1}$ \quad The Abdus Salam International Centre for Theoretical Physics (ICTP), Trieste 34151, Italy; canessae@ictp.it\\
}

\abstract{A comprehensive overview of the recent physics-inspired genome analysis tool, GenomeBits, is 
presented. This is based on traditional signal processing methods such as Discrete Fourier Transform (DFT). 
GenomeBits can be used to extract underlying genomics features from the distribution of nucleotides, and 
can be further used to analyze the mutation patterns in viral genomes. Examples of the main GenomeBits findings 
outlining the intrinsic signal organization of genomics sequences for different SARS-CoV-2 variants along the 
pandemic years 2020-2022 and Monkeypox cases in 2021 are presented to show the usefulness of GenomeBits. 
GenomeBits results for DFT of SARS-CoV-2 genomes in different geographical regions
are discussed together with the GenomeBits analysis of complete genome sequences for
the first coronavirus variants reported: Alpha, Beta, Gamma, Epsilon and Eta. Interesting
features of the Delta and Omicron variants in the form of a unique ‘order-disorder’ 
transition are uncovered from these samples as well as from their cumulative distribution function 
and scatter plots. This class of transitions might reveal the cumulative outcome of mutations on the spike 
protein. A salient feature of GenomeBits is the mapping of the 
nucleotide bases (A,T,C,G) into an alternating spin-like numerical sequence via a series having 
binary (0,1) indicators for each A,T,C,G. This leads to derive a set of 
statistical distribution curves. Furthermore, the quantum-based extension of the GenomeBits 
model to an analogous probability measure is shown to identify properties of genome 
sequences as wavefunctions via a superposition of states. An association of the 
integral of the GenomeBits coding and a binding-like energy can in principle also be established. 
The relevance of these different results in Bioinformatics is analyzed.}


\begin{document}


\section{Introduction}

Bioinformatics methods used to identify genome signatures usually associate some numerical encoding
to strands of symbolic nucleotide letters (A)denine, (C)ytosine, (G)uanine and (T)hymine (or (U)racil
nucleobase in the nucleic acid RNA). This implies that nucleotides are
assigned some fixed numerical values for the purpose of extracting information. As a consequence of this
mapping, some features of the original sequence have been derived which are preserved when the
associated numerical representation is unique. In general, the properties of large data sets
correlate a genome sequence when there is not any mapping function degeneracy.

For example, the well-known binary indicator of Voss~\cite{Vos92} converts a DNA sequence to 
four one-to-one sequences of 0 and 1 to represent the characters A,T,C and G correspondingly. 
Signal processing methods such as discrete Fourier transform (DFT) Power Spectrum and iterative graphical 
representations of DNA have been applied by assigning the numerical sequences (0,1)~\cite{Tun16}. 
The goal is to reconstruct the DNA sequence at any point and to display genome signatures of 
the whole sequence starting from these partial indicators.
This is relevant to compare genomics sequences when only parts of genomics strand are available.
These alignment-free comparison approaches have been extensively investigated in the last decade,
and most recently for SARS-CoV-2 sequences~\cite{Tho22}.

Motivated by these simple numerical (0,1) representations of biological sequences, and 
in response to the urgent need of research on SARS-CoV-2 during the global pandemic years 2020-2022, 
we introduced a new numerical mapping algorithm, named GenomeBits, to unreveal intrinsic signals 
from coronavirus (hCoV-19) sequences in FASTA format. In a series of papers written during the pandemic,
interesting GenomeBits findings for different pathogen variants were reported in several Refs.~\cite{Can1,Can2,Can3,Can4}.
These investigations are now summarized in this overview with new examples to demonstrate the utility of
this alternative Bioinformatics method. 

The significance of GenomeBits is to surpass some limitations of other statistical procedures 
which seldom support a focused analysis at each single nucleotide level. Alternative
frameworks for handling genomics information in the strands of DNA and RNA molecules from pattern matching
consider symbolic codons (i.e., triplets of nucleotides) sometimes through manual examination.
State-of-the-art methods for genomics analysis to identify variations and characteristic profiles still 
need algorithms to compare complete genome sequences quickly, accurately and efficiently~\cite{Als22}. 
In this light, GenomeBits might become another feather in Bioinformatics toolkits for genome characterization.
The GenomeBits approach has in practice a twofold advantage: first, it allows to explore possible short range trends 
and local features of complex adenovirus sequences. The discovery of genomic signals at "small" scales
--say of 30,000 base position (bp), is potentially significant to search relevant genomic signals 
and detect general trends and basic properties. The second salient feature of the GenomeBits approach is that 
it provides insights for each nucleotide bases A,C,G and T separately, adds capabilities for fast 
sequence taxonomy and easy data visualization. Our hope is to encourage researchers to develop and apply
physics-inspired analysis tools, like GenomeBits, based on traditional signal processing methods
to extract underlying genomics features from the distribution of nucleotides and to compare mutation
patterns in viral genomes.

The article is structured as follows. Section 2 describes in great detail
the GenomeBits method. We discussed how the nucleotide bases A,T,C and G are 
mapped as a spin-like numerical sequence via a finite alternating sum series having
a distribution of (0,1) indicators for each A,T,C and G. 
This expansion of the genome sequences from 1D to 4D leads to evaluate computationally inexpensive 
statistical distribution curves. Section 3 displays the results of using the Fourier transform 
of SARS-CoV-2 genomes in different geographical regions.
These include GenomeBits analysis for complete genome sequences for 
the first coronavirus variants reported: Alpha, Beta, Gamma, Epsilon and Eta. Interesting 
features of the other coronavirus variants Delta and Omicron in the form of a
unique ‘order-disorder’ transition is uncovered from these samples as well as their
cumulative distribution function (CDF) and scatter plots.
Section 4 introduces the quantum-inspired GenomeBits model extension.
Such analogous probability measure allows to identify emergent properties of genome sequences 
in the form of wavefunctions via a superposition of states. 
The final section (Section 5) concludes this review with a possible 
association of the integral of the GenomeBits coding and a binding-like energy,
and proposes some future directions for research.

\section{GenomeBits mapping}

The GenomeBits mapping to retrieve special patterns of complete genome 
sequences assigns alternating sum series having multiple (0,1)
values for the nucleotide variables $\alpha = A,C,T,G$ of genome 
sequences~\cite{Can1}. The mapping of letter sequences into binary values is not new 
as already mentioned~\cite{Vos92}. The novelty in the numerical encoding of GenomeBits 
lies in the use of positive and negative signs $(-1)^{k-1}$ in the $k$-base sums.
The analysis of genomics sequencing throughout finite alternating sums allows 
to extract distinctive features at each nucleotide base. From a statistics  point of view, 
these alternating series correspond to a time series of discrete values.

The arithmetic progression in question carries in fact alternating terms, and a finite 
positive moment of independently distributed variables $X_{k,\alpha}$,
as follows 
\begin{equation}
\label{eq:eq1}
     E_{\alpha,N}(X) = \sum_{k=1}^N (-1)^{k-1}X_{\alpha,k} \; .
\end{equation}
The individual terms $X_{k}$ are associated to 0 or 1 values according to their 
base position along the genome sequences of length $N$. They are assumed to satisfy the relation
\begin{equation}
     X_{\alpha,k=N} = | E_{\alpha,N}(X) - E_{\alpha,N-1}(X) | \; .          
\end{equation}
In this GenomeBits mapping the binary indicator for the sequences assumes alternating positive 
and negative signs. The ($\pm$) signs are assigned sequentially starting 
with $+1$ at $k=1$ as in the example of Table~\ref{table1}. If a term $X_{\alpha,k}$ 
is positive or zero ($1$ or $0$) at a given nucleotide bp $k$, then the next 
$X_{\alpha,k+1}$ term can be negative or zero again ($-1$ or $0$). This selection 
is inspired by the discrete physics Ising spin model in which variables 
can be in one or more states with spin up or spin down in a lattice
(representing magnetic dipole moments of atomic spins). 
It can also be interpreted in terms of a Balanced Ternary Logic which 
uses the three digits -1, 0, and +1. Balanced ternary has many applications
including the unit of quantum information realized by a 3-level quantum system (qutrit or quantum trit), 
that may be in a superposition of mutually orthogonal quantum states --just as the Qubit.

A GenomeBits mapping example of the hCoV-19 genome fragment GTATACTGCTGC (having $12$ nucleotides) 
converted to the alternating (0,1) array via Eq~(\ref{eq:eq1}) is the following

\begin{table}[H]
\begin{adjustwidth}{-2.8cm}{0cm}
\caption{{\bf GenomeBits mapping} via Eq~(\ref{eq:eq1}) for $N=12$.\label{table1}}
\newcolumntype{C}{>{\centering\arraybackslash}X}
\begin{tabularx}{1.2\textwidth}{|C| c c c c c c c c c c c c |C|} 
\toprule
Base position $k$  & 1 & 2 & 3 & 4 & 5 & 6 & 7 & 8 & 9 & 10 & 11 & 12 &  GenomeBits sums \\
\midrule
Sequence (strand)  & {\bf G} & {\bf T} & {\bf A} & {\bf T} & {\bf A} & {\bf C} & {\bf T} & {\bf G} & {\bf C} & {\bf T} & {\bf G} & {\bf C} &  $\sum_{k=1}^{12} (-1)^{k-1}X_{\alpha,k}$   \\ \hline 
$(-1)^{k-1}X_{\alpha=A,k}$ & 0 & 0  & +1 & 0  & +1  & 0  & 0  & 0  & 0  & 0  & 0  & 0  &  +2  \\ \hline
$(-1)^{k-1}X_{\alpha=C,k}$ & 0  & 0  & 0  & 0  & 0 & -1  & 0  & 0  & +1  & 0 & 0  & -1 & -1  \\ \hline
$(-1)^{k-1}X_{\alpha=G,k}$ & +1  & 0 & 0  & 0  & 0  & 0  & 0 & -1  & 0  & 0  & +1  & 0  &  +1  \\ \hline
$(-1)^{k-1}X_{\alpha=T,k}$ & 0  & -1  & 0  & -1 & 0  & 0 & +1  & 0 & 0 & -1  & 0 & 0  & -2  \\ \hline
\bottomrule
\end{tabularx}
\end{adjustwidth}
\end{table}

On the other hand, a quantum-based extension of our GenomeBits approach to retrieve properties 
of genome sequences converted to 0 and 1 outcomes can also be derived from the above class of
series~\cite{Can4}. In this way, GenomeBits can become a
measurement theory and not a model of quantum physical processes at the atomic level.
This "analogous" extension to (real and imaginary parts of longitudinal) wavefunctions 
represents a mathematical description of an isolated, "analogous" quantum system where the 
wavefunction {\emph vs.} the nucleotide bases present features of sound waves. 
This approach leads to reveal novel measures of the genome evolution 
at nucleotide levels during genome mutations over $N$ intervals. 

Let us consider Eq.(\ref{eq:eq1}) as the resulting wave generated 
by a certain superposition of multiple discrete wavefunctions, {\em i.e.}, 
$\phi(X_{\alpha,N}) \rightleftharpoons \psi_{n}(X_{\alpha,N})$ in a medium. 
In polar form, we then extend the GenomeBits approach to write
\begin{eqnarray}\label{eq:psi}
\psi_{n}(X_{\alpha,k}) & \equiv & A\; (-1)^{k-1} | \Phi(X_{\alpha,k}) - \Phi(X_{\alpha,k-1}) |       
                          \; \exp\left\{\frac{n\pi i}{\lambda_{_{N}}}\Phi(X_{\alpha,k})\right\} \; ,  \nonumber \\
                           &    =    &   A\; (-1)^{k-1}X_{\alpha,k}       
                          \; \exp\left\{\frac{n\pi i}{\lambda_{_{N}}}E_{\alpha,k}(X)\right\} \; , 
\end{eqnarray}
with $A$ a real constant and $n = 1,2 \cdots$ 
These wavefunctions are complex functions in general and the displacement of the wave is a function of 
the $k$-bp. The normalization condition via the complex conjugate 
$\sum_{k=1}^N |\psi_{n}(X_{\alpha,k})|^{2} = \sum_{k=1}^N \psi_{n}(X_{\alpha,k})\psi_{n}^{*}(X_{\alpha,k}) = 1$ 
leads to
\begin{equation}
A^{2}\sum_{k=1}^N |(-1)^{k-1}X_{\alpha,k}|^{2} = A^{2}\sum_{k=1}^N X_{\alpha,k}^{2} = 1 \; .
\end{equation}
Since $X_{\alpha,k}$ takes (0,1) values only, the amplitude then satisfies
\begin{equation}\label{eq:a}
A = \frac{1}{\pm\sqrt{N_{_{+1}}}} \; ,
\end{equation}
where we denote by $N_{_{+1}}$ the total number of 1's found in a complete $N = N_{_{0}} + N_{_{+1}}$ 
sequence within each species $\alpha$. 
The maximum real value for the alternating sum of binary sequences of Eq.(\ref{eq:eq1})
can be obtained by the choice
\begin{equation}
\lambda_{_{N}} \equiv \sum_{j=1}^N \phi(X_{\alpha,j}) = \sum_{k=1}^N (-1)^{k-1}X_{\alpha,k} \equiv E_{\alpha,N}(X) = \Phi(X_{\alpha,N})\; ,
\end{equation}
From~Eq.(\ref{eq:psi}) and Euler's identity, we note that for $k \rightarrow N$ and $\forall n \geq 1$,
the Cartesian form of $\psi$ oscillates and decreases for large numbers of 1's, namely
\begin{equation}\label{eq:psiN}
\psi_{n}(X_{\alpha,N}) = \left(\frac{1}{\pm\sqrt{N_{_{+1}}}}\right) (-1)^{N-1} X_{\alpha,N} \; [\cos(n\pi) + i \;\sin(n\pi)] 
                              =  \frac{(-1)^{n}}{\pm\sqrt{N_{_{+1}}}}\; \phi(X_{\alpha,N}) \; .
\end{equation}
Hence, the GenomeBits~Eq.(\ref{eq:eq1}) 
for complete genomics strands can be seen as a steady state wave created by some non-zero 
complex wavefunctions $\psi$. The wavefunction changes with $k \ne 0$ having maximum density probability 
$1/N_{_{+1}}$ at each $n$-value.

\section{Discussion}

The complete genome sequence of the contagious SARS-CoV-2 coronavirus in humans were first 
reported by Chinese scientists in early 2020~\cite{Can1}. To this date, there are thousands 
sequence sources openly available through the well-known GISAID Initiative ({\tt www.gisaid.org}) 
that allowed comparisons and classifications of species of different emerging 
coronavirus variants and genome mutations from around the globe during all the pandemic years 2020-2022.
These data were extremely useful for genomics surveillance and, today, they 
form a rich source of information for the development of new theories that may predict 
and catalog this class of genomics strands. For almost half of its genome, the coronavirus 
presented a unique lineage with only few relationships to other known viruses, in particular
in the spike region encoding the S-protein.

\subsection{DFT Power Spectrum}

Using GISAID data, we presented in~\cite{Can1} the GenomeBits signal analysis of 
Eq.(\ref{eq:eq1}) for the complete genome sequenced for coronavirus variants B.1.1.7 (Alpha), 
B.1.135 (Beta), P1 (Gamma), B.1.429–B.1.427 (Epsilon) and B.1.525 (Eta), with $N$ nucleotides 
on the order of 30k bp in length.
These samples were taken from broad geographical regions and were collected during different
periods of the pandemic starting November 2020 up to March 2021. In~\cite{Can1}
we discussed how the GenomeBits method provides additional information to conventional
Similarity comparisons via alignment methods and DFT Power Spectrum approaches. To obtain 
these results, a simple Graphics User Interface (GUI) was implemented which can be downloaded from 
Github~\cite{bib-1h}. The Linux-based GUI runs under Ubuntu O.S. and requires little processing time 
for the analysis of complete genomics data. The GenomeBits GUI considers samples with A,C,T,G 
sequences corresponding to genomics sequence data from given variants and different Countries.
Sequences uncompleted with codification errors (for example '$NNNN$' letters) 
are not considered.

\begin{figure}[H]
\begin{adjustwidth}{-2.8cm}{0cm}
\includegraphics[width=8.0 cm]{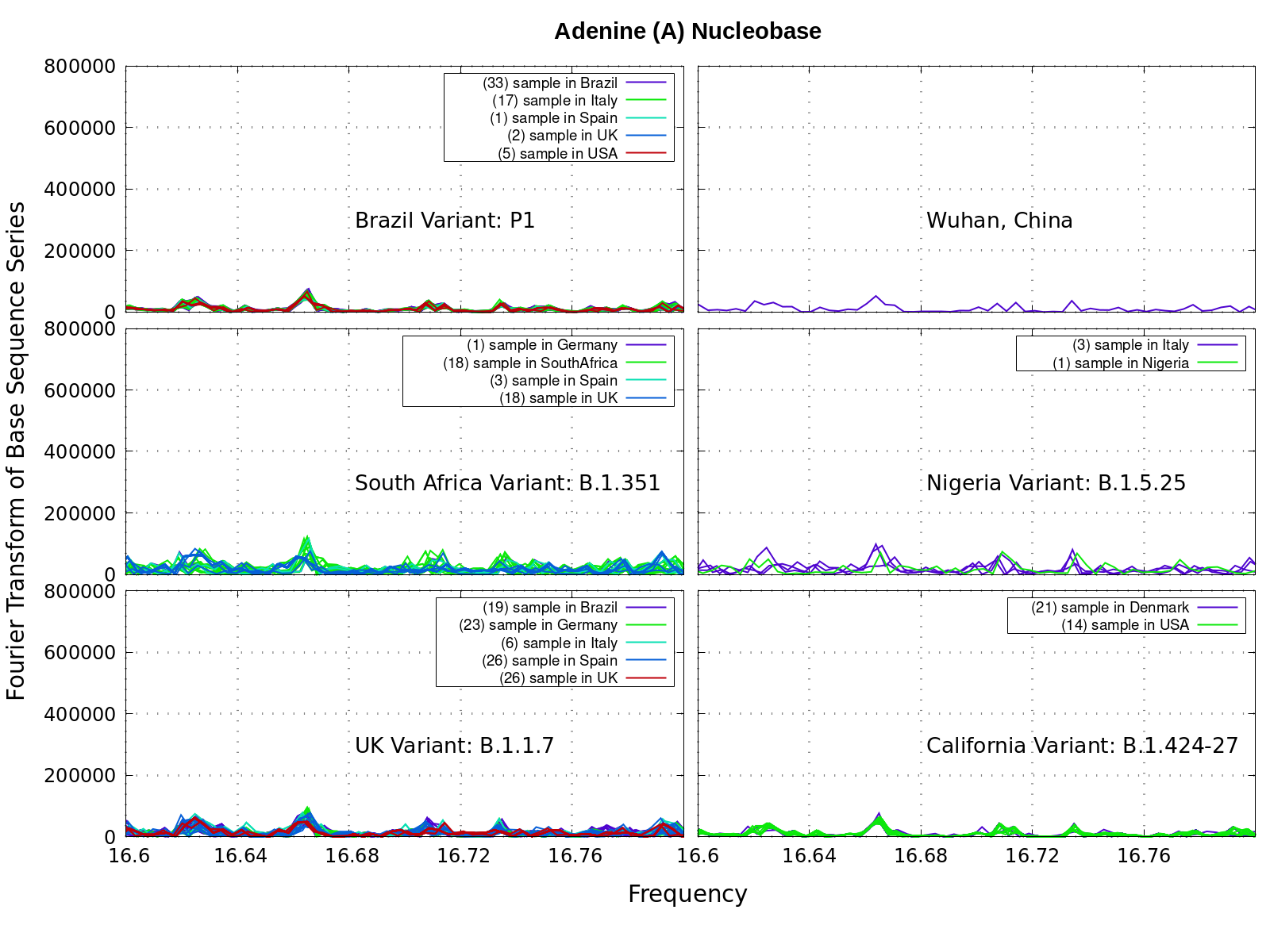}
\includegraphics[width=8.0 cm]{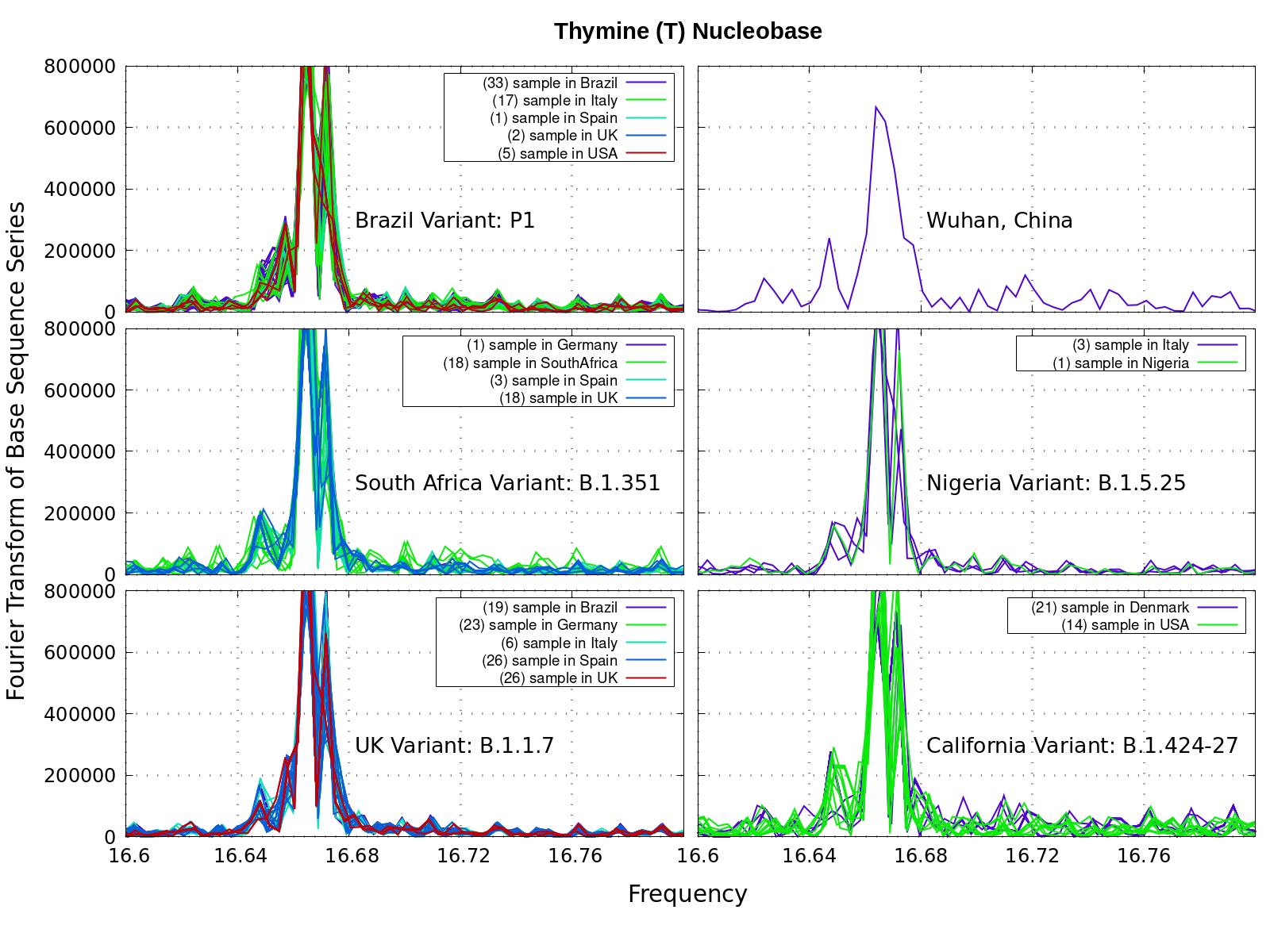} \\
\includegraphics[width=8.0 cm]{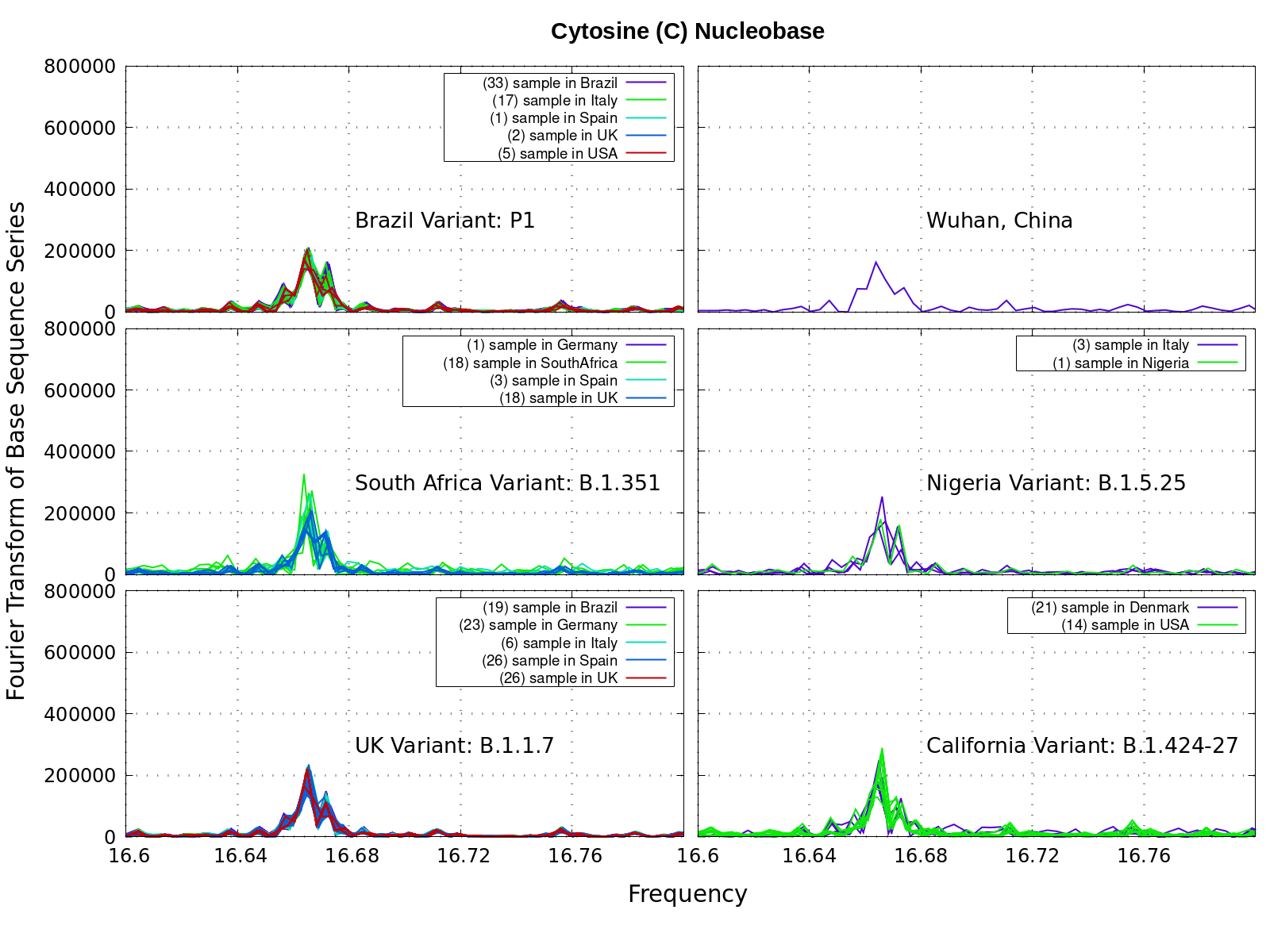}
\includegraphics[width=8.0 cm]{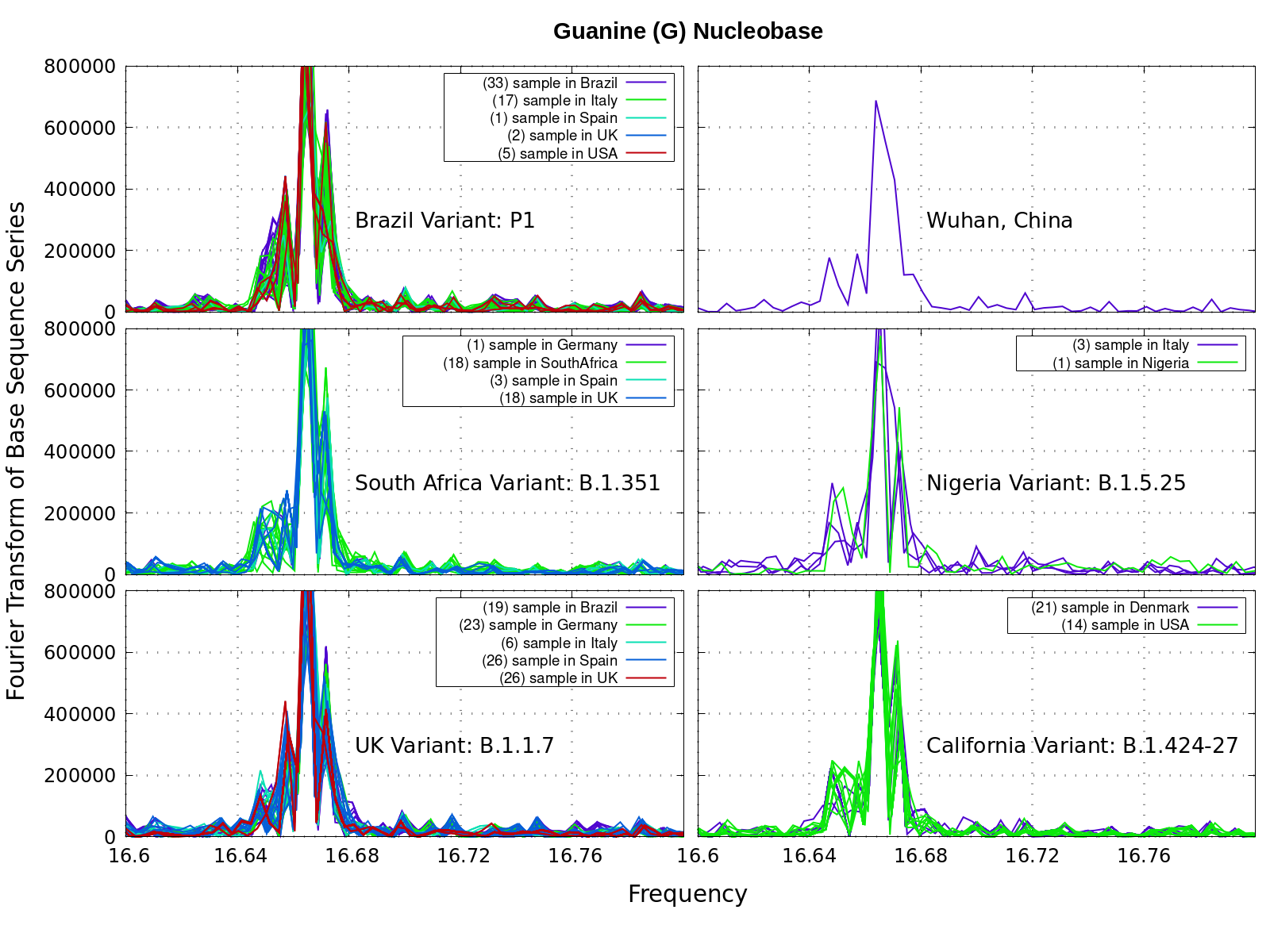} \\
\caption{
Discrete Fourier transform of the alternating sum series of Eq.(\ref{eq:eq1})
for nucleotide bases according to its progression $N$ along
different samples of coronavirus genome available from different countries.
\label{fig1}}
\end{adjustwidth}
\end{figure}

As mentioned, GenomeBits can be used to investigate the conventional 
DFT Power Spectrum to identify base periodicity properties of genome 
sequences and classification variants during pandemic. In general the Power 
Spectrum of the microorganism sequence is usually considered as the sum of the partial spectra: 
$\sum_{\alpha} |S_{\alpha}(f)|^{2} = (1/N^{2})\sum_{\alpha}^{N}|X_{\alpha,k}\; exp(2\pi ifk)|^{2}$,
with discrete frequencies $f=1/N,\; 2/N\; \cdots$ 
The DFT maps genome sequences into four (0–1) indicators 
and evaluates them in frequency domain. In our method, DFT can provide some insights for single 
nucleotide bases A,C,G and T to characterize virus variants as illustrated in Fig.\ref{fig1}
for nucleotide bases from different variants and countries (same as in~\cite{Can1}).

As can be seen in the plots, the DFT of the alternating sum series of Eq.(\ref{eq:eq1})
reveals partial peaked structures appearing at a ‘frequency’ equal to 16,66 implying for the
alternating sums a $50/3$ characteristic periodicity each above noise level. We note that larger
peaks appear in the nucleotides of the complementary strands T and G only. This interesting
result can be detected only in frequency domain.  In~\cite{Can1}, it was shown that their total
sum $|S_{A}|^{2}+|S_{T}|^{2}+|S_{C}|^{2}+|S_{G}|^{2}$  has a peak at around frequency
$33.33$ for all variants considered. This is a unique, distinctive pattern retrieved
from the intrinsic data organization of genome sequences according to their progression
along the nucleotide bp.

\subsection{‘Order-disorder’ transition}

\begin{figure}[H]
\begin{adjustwidth}{-2.8cm}{0cm}
\includegraphics[width=16.0cm]{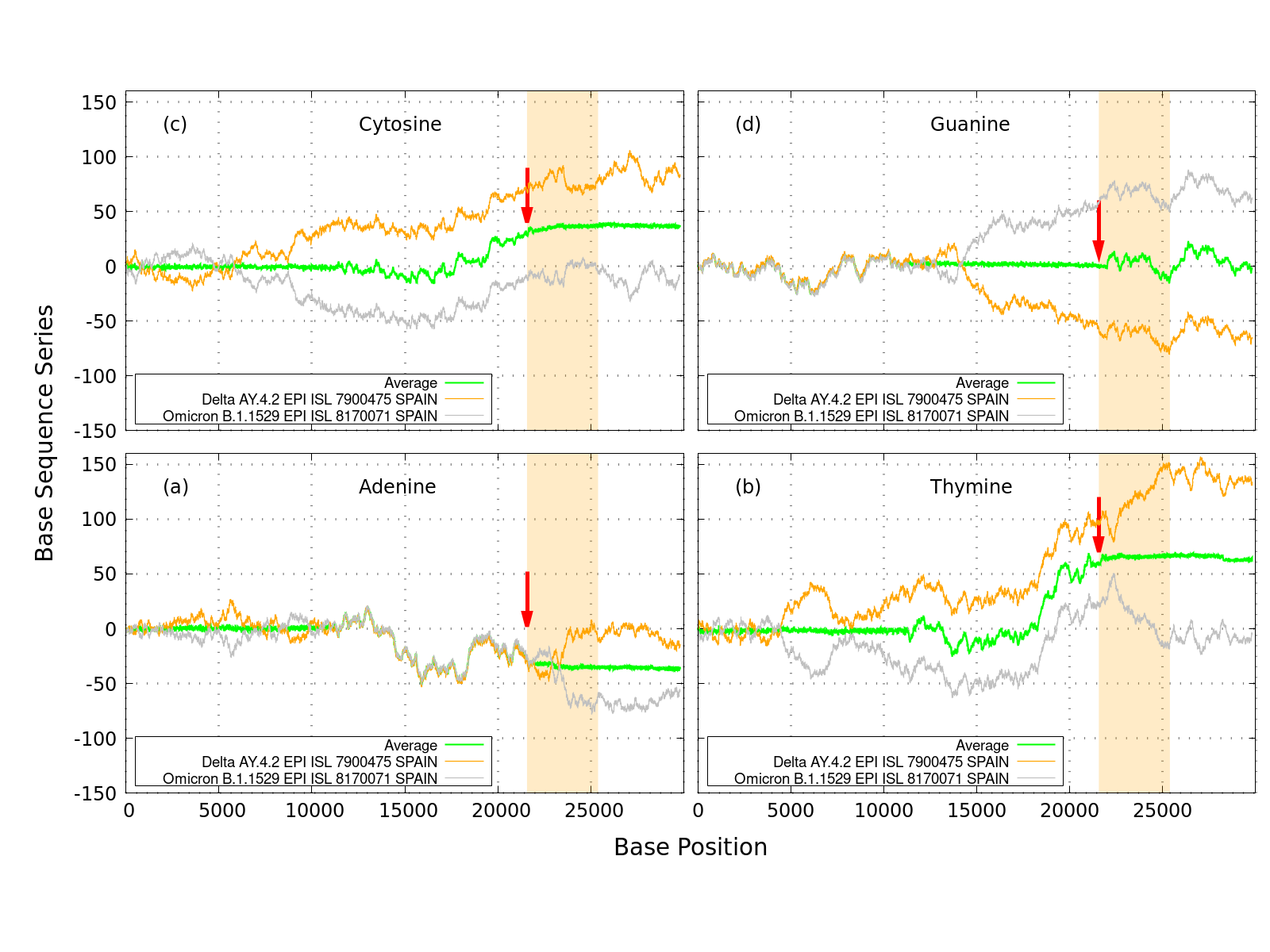}
\caption{
GenomeBits sum series: Delta (in orange) and Omicron (in gray) variant imprints
for coronavirus samples from Spain. Arrows show a ‘disordered’ (peaked)
to ‘ordered’ (constant) transition around the S-spike region of the SARS-CoV-2 Wuhan-Hu-1
sequence shown in clear red. At the arrow-pointed region of Guanine
in subfigure (d), larger ‘disordered’ structures are found inside the S-spike
region in contrast with those for A,C, and T (i.e., subfigures (a),(b) and (c))
which appear outside the S-spike region displaying a more ‘ordered’ transition.
\label{fig2}}
\end{adjustwidth}
\end{figure}

The GenomeBits Eq.(\ref{eq:eq1}) can also reveal another 
interesting underlying genomics feature of coronavirus variants, namely, an ‘order-disorder’ 
transition~\cite{Can2}. In Fig.~\ref{fig2}, we show results for the sequences 
of the coronavirus variants AY.4.2 Delta and B.1.1529 Omicron 
samples from Spain together with their average values as indicated.
The Delta variant was first detected in India at the end of 2020 and then
in South Africa in late 2021 to then become dominant world-wide.
delta and Omicron variants had common mutations in the building blocks that conform 
the spike protein (responsible for the pathogen penetration in a human cell).
Omicron caused less severe Covid-19 than Delta variant during the pandemic waves. 
Nature selected Omega mutations which seemed to replicate more efficiently.

On the left of Fig.~\ref{fig2} results for the pair nucleotides A,C and on the right the
complementary nucleotides T,G are shown. In the figure we found regions where the 
alternating sums from Delta data (in orange) mirror those of the Omicron (in gray).
This behavior is highlighted by the green values after averaging both curves. 
The regions having low data noise (or rather constant averaged values), indicate 
coding correspondence between variants specially around the coronavirus S-spike region.
These results reveal a sort of ‘disordered’ (or peaked) to ‘ordered’ (or constant) transition indicated 
by the red arrows for each single nucleotide level. This may reflect the polymorphism in the 
genomics variations by only assigning alternating (0,1) values to symbolic nucleotide characters. 

It is important to observe the distinctive Guanine patterns for Delta and Omicron variants of coronavirus 
samples from Spain shown on the top right of Fig.~\ref{fig2}(d). The Guanine ‘ordered-to-disordered’ GenomeBits 
transition appears to be different from the other three nucleic acid bases A,C and T. At the arrow-pointed 
region, the Guanine transition appears to have been inverted. Larger (‘disordered’) peaked structures are 
found inside the S-spike region in contrast with those for A,C, and T which appear outside the S-spike region 
displaying a more ‘ordered’ (or constant) transition (i.e., Figs.~\ref{fig2}(a)(b)(c)). This class of profile 
might reveal the cumulative outcome of mutations on spike protein from Delta to Omicron coronavirus variants 
by some A,C, or T > G substitutions or vice-versa. Significant differences in the spike region can impart 
these unique properties found for Guanine. The E484 SARS-CoV-2 Spike protein mutation corresponds to 
a nucleotide substitution from Guanine (as in the original Wuhan-Hu-1 isolate) to Adenine (Beta and Gamma 
variants) or Cytosine (Kappa variants) (see \cite{Zha22}). Alongside, the observed ‘disordered’ transitions 
in the non-spike regions for A,C,T curves (displayed around 10,000-20,000 base positions) could also be 
responsible elements behind single-nucleotide mutations, which might led to properties including growth 
advantages, transmission rate and immune evasion, among others.

\subsection{Statistical imprints}

The evolution of other infectious disease relating Monkeypox virus (MPXV) also alerted
humanity in 2022. Soon after, the genome sequences for MPVX were analyzed using the GenomeBits 
method~\cite{Can3}. In that study, histograms, empirical and theoretical cumulative distribution curves 
and the resulting scatter plots for the base nucleotides $A-C$ versus their complementary base nucleotides 
$T-G$ for MPVX were reported. In the following we extend these statistical investigations and apply them 
anew to the signal analysis of typical hCoV-19 genome sequences of lineage B.1.1.529 Omicron from USA
(ID EPI\_ISL\_7887528 and EPI\_ISL\_7887531).

\begin{figure}[H]
\begin{adjustwidth}{-2.8cm}{0cm}
\includegraphics[width=8.0 cm]{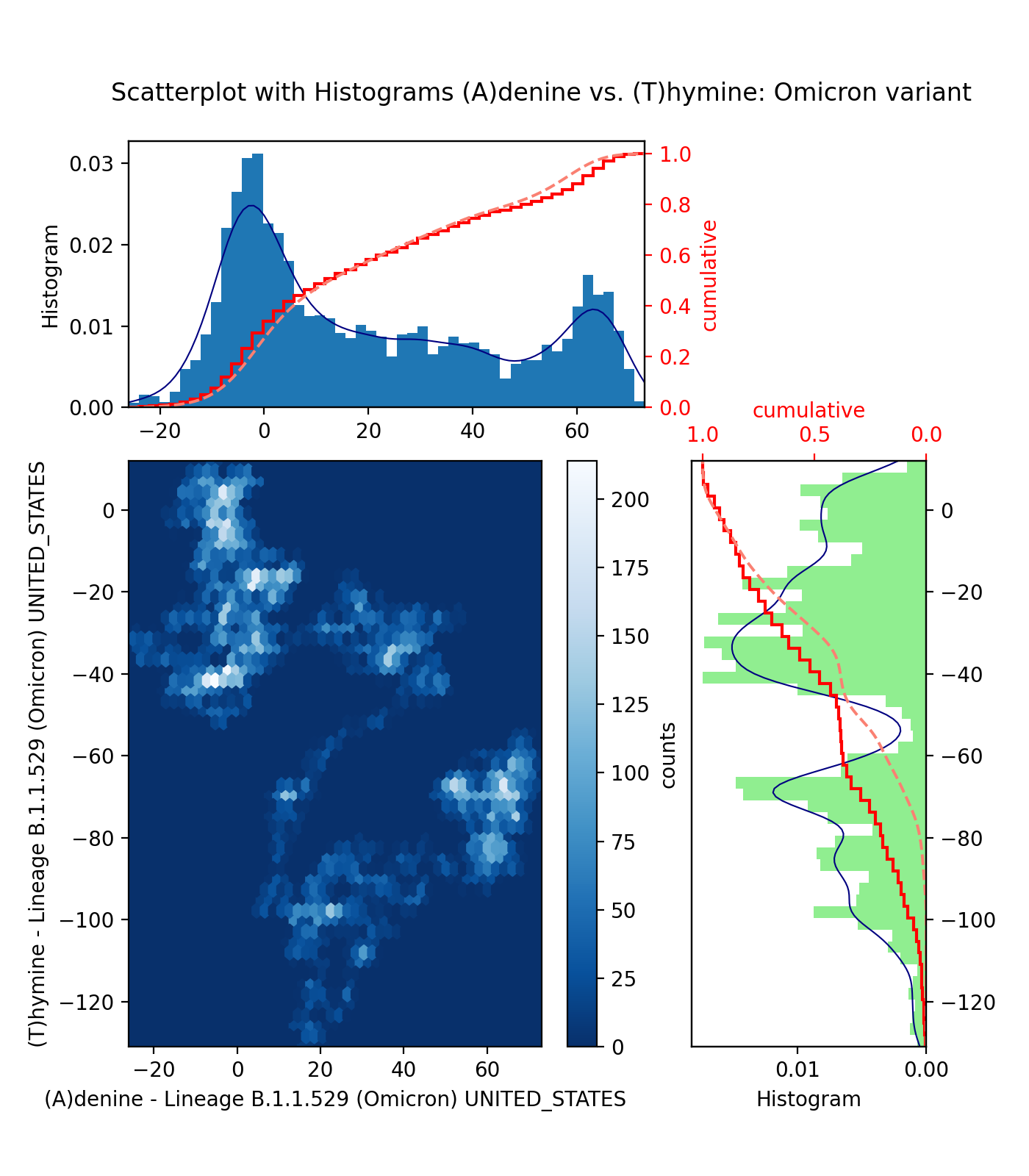} 
\includegraphics[width=8.0 cm]{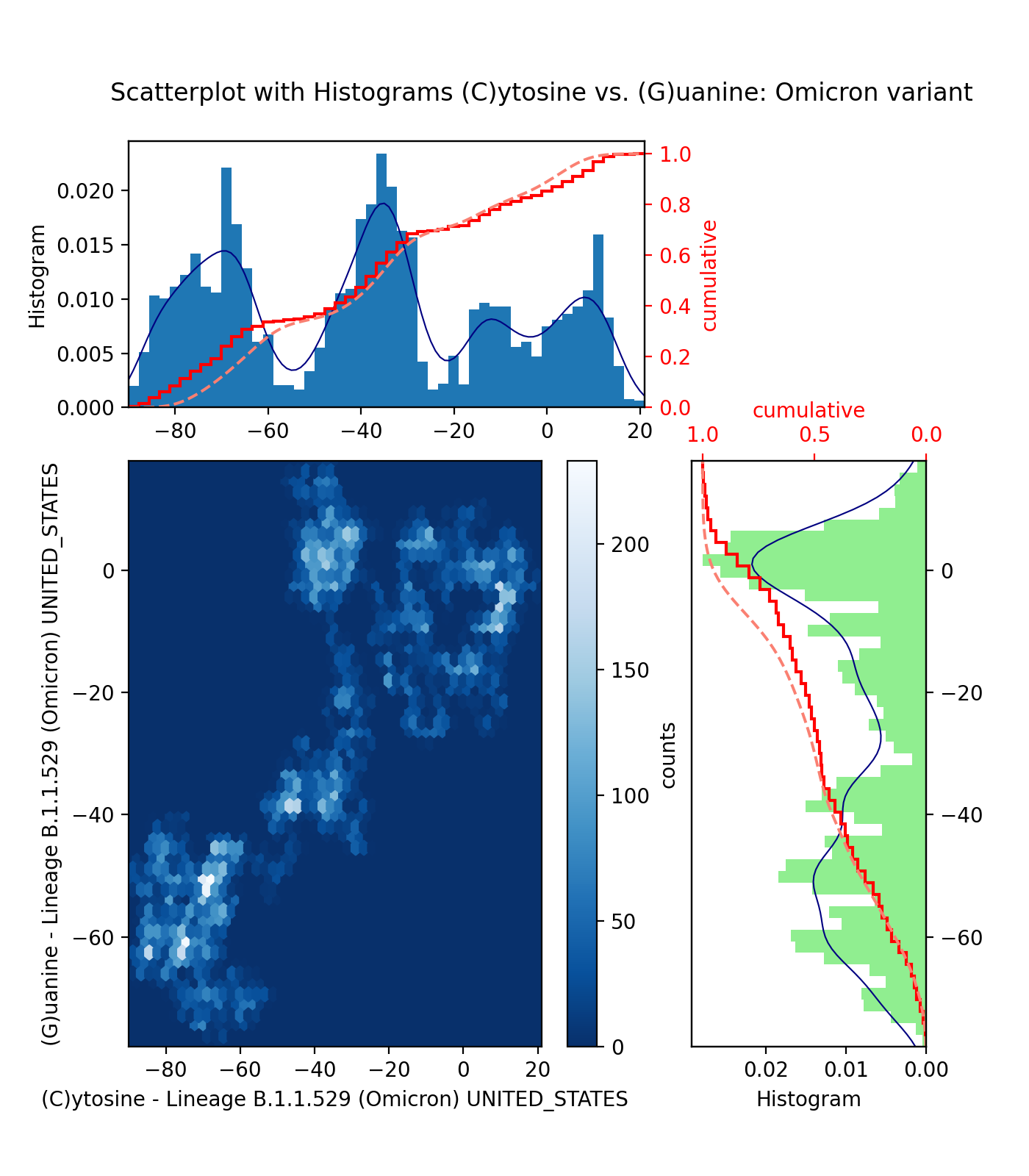} 
\caption{
Scatter plot of GenomeBits Adenine and Thymine curves with histograms from the hCoV-19 
genome sequences of lineage B.1.1.529 Omicron from USA (ID EPI\_ISL\_7887528 and EPI\_ISL\_7887531).
\label{fig3}}
\end{adjustwidth}
\end{figure}

In Fig.~\ref{fig3}, we show the histogram curve (i.e., number                                     
of data points in a given bin), a standard CDF and 
the scatter plot obtained from two different histogram data sets of equally sized bins.
In the figure, the blue and green bar-graph are the histograms and the red full line and 
red dotted lines are empirical and theoretical CDF curves, respectively. 
In the computations 50 bins are used, and the blue line fits
have been obtained via a Gaussian distribution.

The resulting scatter diagrams for $A-C$ versus $T-G$ base nucleotides allow to highlight 
correlations and anomalies at certain colored contact points between GenomeBits sequences 
of complementary nucleotide bases. The clearer colors of the hexagon markers add a new 
dimension to the study of patoghens. In fact, the plotted patterns allow for visual 
comparisons. Markers show how the mapped 
binary data is stratified along the sequences and how correlated points fall along distinct 
shapes. The patterns found are unequally distributed in the scatter plot and these do not occur 
randomly.

In this overview, we considered the distinctive patterns of complete 
full letter sequence of all the genes of the coronavirus genome
for single strand as available from, and reported within, the GISAID Initiative for the 
building blocks of nucleic acid coding sequence Adenine, Cytosine, Guanine and Thymine.
In RNA, Thymine is replaced by Uracil because of their six-member single-ring structural similarity 
referred to as pyrimidines. The chemical structures of Uracil and Thymine, composed of carbon and 
nitrogen atoms, are very similar. The presence of Thymine in a DNA strand in actively 
dividing cells is more stable thermodinamically and improves the efficiency of DNA replication 
when compared to Uracil in RNA. The complementary base of both Uracil and Thymine is Adenine.
During the pandemic, it was observed that SARS-CoV-2 genomes contain more Uracil than any other nucleotide.
Substitution of nucleotides to Uracil was highest among the non-synonymous mutations observed. 

From these simple statistical imprints, we believe that the GenomeBits method can help 
to shed light behind the behavior of infectious diseases by focusing on single-nucleotide 
structures. Structures which are different on each mutation of a virus.

\section{Wave-like features}

Acoustic-like sounds within a biological context can help to identify 
trends in genome sequences and characterize new properties. For example, sonification algorithms 
based on biological rules for DNA sequences have used codons to generate audio strings 
representative of synthesized RNA during transcription~\cite{Plai21}.
Another possibility to identify emergent properties of genome sequences in the form of complex
wavefunctions can be obtained by Eq.(\ref{eq:psi}). 
This wavefunction becomes a mathematical description of an analogous quantum system.

In this context, it is worthy to note that at the initial base position $k = 1$, the following relations follow
from our previous equations
\begin{eqnarray}\label{eq:relationsk1}
|\psi_{n}(X_{\alpha,1})|^{2}  &  =  &  \psi_{n}(X_{\alpha,1}) \; \psi_{n}^{*}(X_{\alpha,1})  \; ,  \nonumber \\
     &  =  & a_{1}^{2} \left( \Psi_{n}^{(a)}(X_{\alpha,1}) \right)^{2} +
                  b_{1}^{2} \left( \Psi_{n}^{(b)}(X_{\alpha,1}) \right)^{2}  \; ,  \nonumber \\
     &  = &  \frac{1}{N_{_{+1}}}   \; X_{\alpha,1}^{2}  \; ,
\end{eqnarray}

and also

\begin{eqnarray}\label{eq:realimagk1}
  b_{1}\Psi_{n}^{(b)}(X_{\alpha,1})  & = & 
      a_{1}\Psi_{n}^{(a)}(X_{\alpha,1}) \; tan \left( \frac{n\pi}{\lambda_{_{N}}} X_{\alpha,1} \right)
    =  \frac{1}{\pm\sqrt{N_{_{+1}}}}   \; X_{\alpha,1} \; sin \left( \frac{n\pi}{\lambda_{_{N}}} X_{\alpha,1} \right) \; ,  \nonumber \\
  a_{1}\Psi_{n}^{(a)}(X_{\alpha,1})  & = & 
     \frac{1}{\pm\sqrt{N_{_{+1}}}}   \; X_{\alpha,1} \; cos \left( \frac{n\pi}{\lambda_{_{N}}} X_{\alpha,1} \right)  \; .
\end{eqnarray}

This means that the wavefunction $\psi_{n}$ can be, in principle, further divided in two
wavefunctions $\Psi_{n}^{(a)}$ and $\Psi_{n}^{(b)}$. This may lead to an analysis of spatial stereo
sonification from genome sequences.

The applicability of this approach to sound waves from complete genome sequences of coronavirus 
pathogens was discussed in~\cite{Can4}. At the level of nucleotide ordering for different genome
mutations, this description of the genome dynamics revealed interesting results. As show next, 
the real (and also imaginary) parts of the longitudinal wavefunction vs. the nucleotide bases 
display characteristic features of sound waves starting with inputs of ones and zeros only. 

\begin{figure}[H]
\begin{center}
\includegraphics[width=9.0 cm]{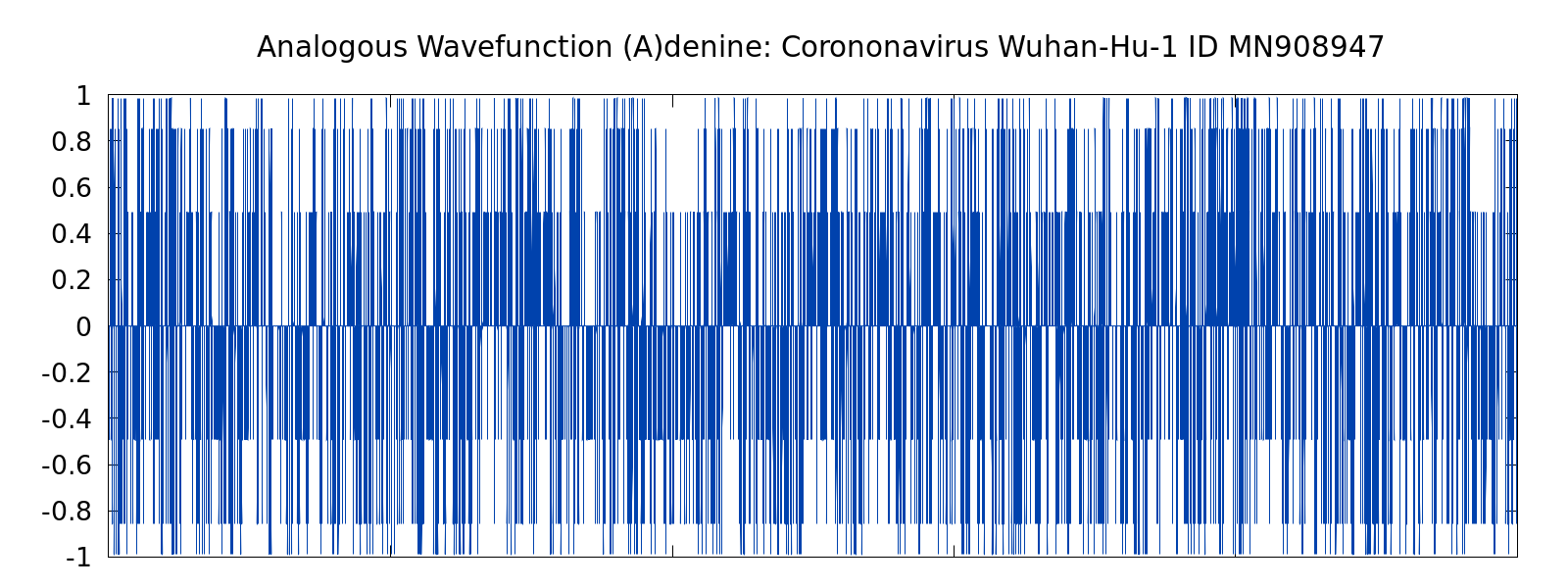} \\
\includegraphics[width=9.0 cm]{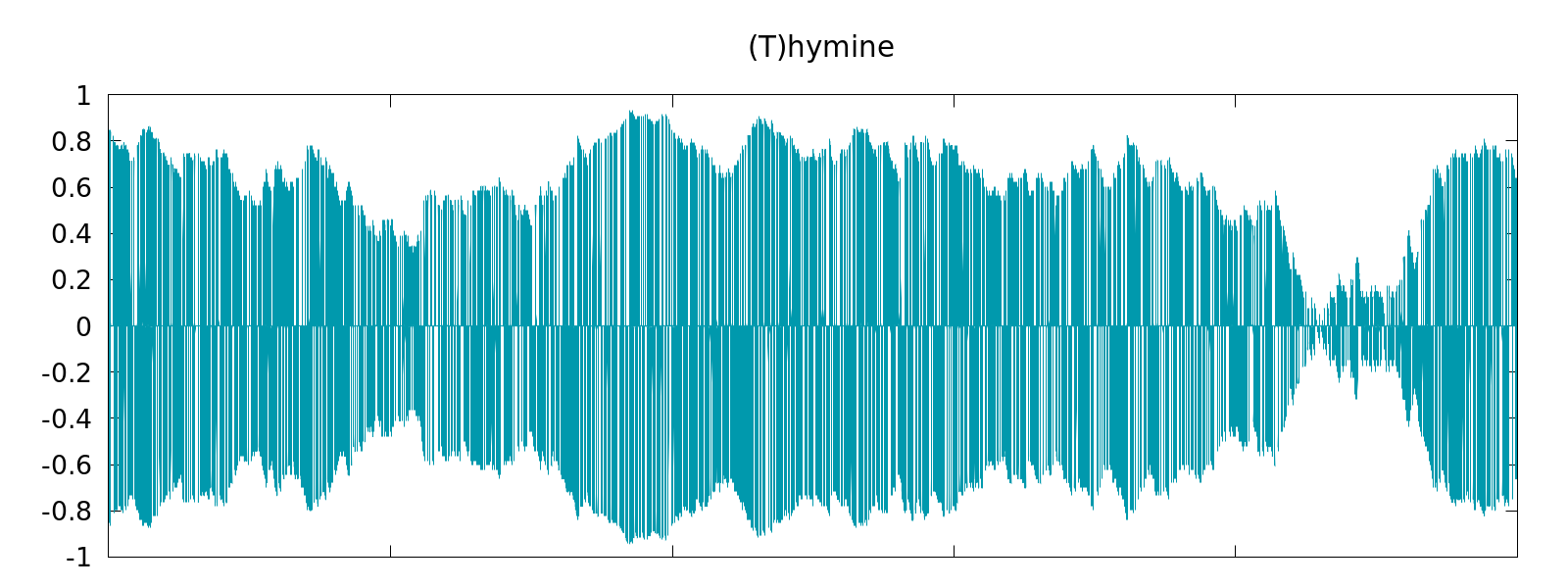} \\
\includegraphics[width=9.0 cm]{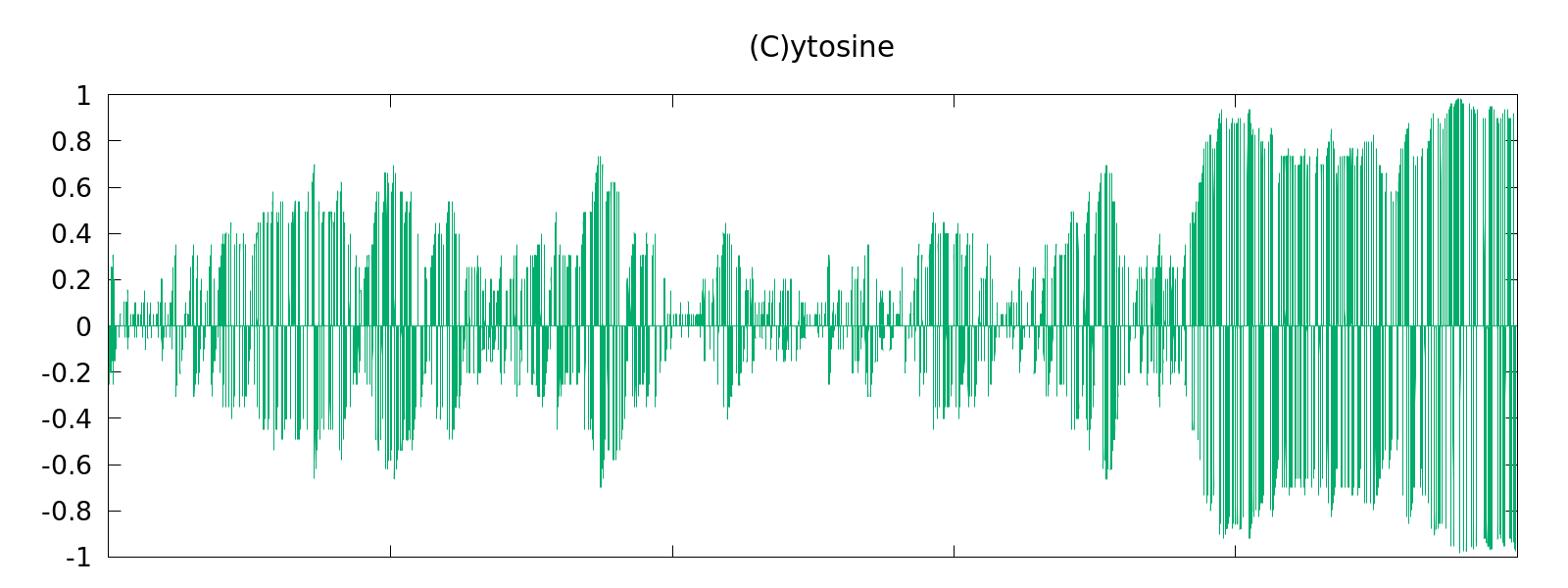} \\
\includegraphics[width=9.0 cm]{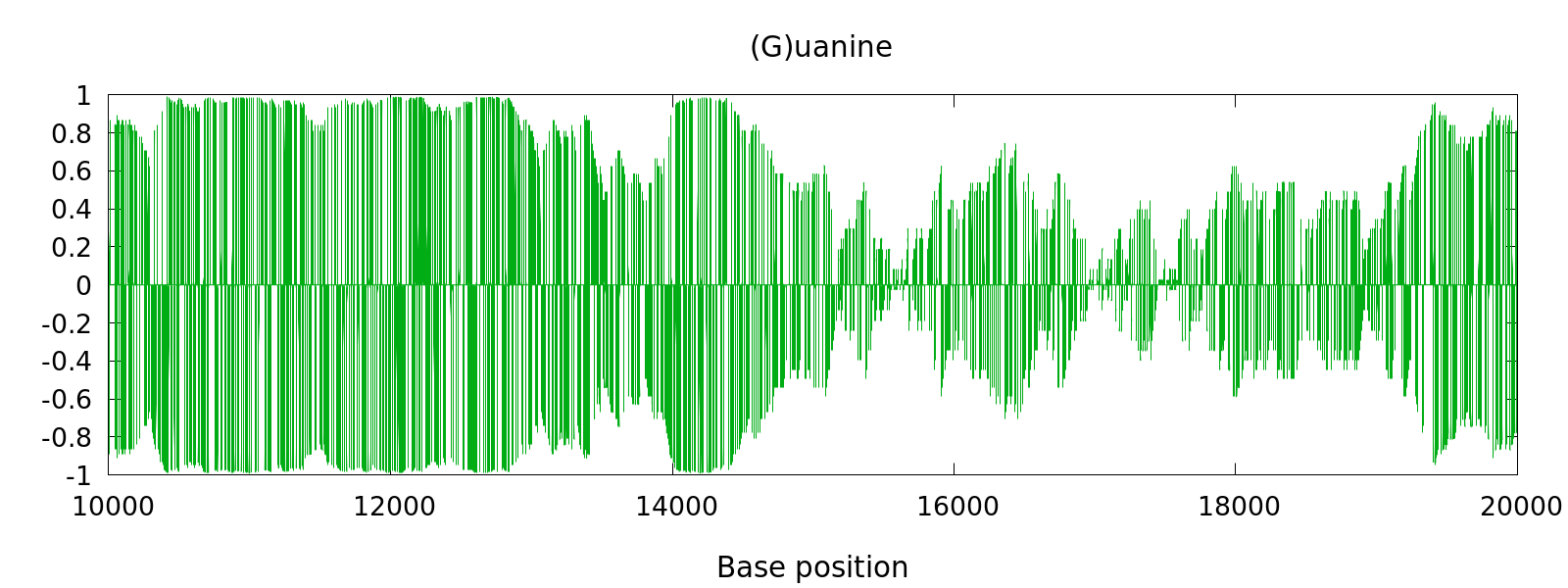}
\caption{
Real part of the wavefunction in $\psi_{n}$ for $n=1$ derived by GenomeBits method.
\label{fig4}}
\end{center}
\end{figure}

As shown by the illustrative results in Fig.~\ref{fig4}, the real spectrum of $\psi_{n=1}$ 
from Eq.(\ref{eq:psiN}) vs. the nucleotide base positions of the 2020 coronavirus Wuhan-Hu-1,
China (ID MN908947) present typical features of sound waves. 
At any point different from zero, binary projections at the level of nucleotides display 
oscillatory patterns in correspondence with the intrinsic gene organization.
This behavior is a consequence of factorizing the discrete wavefunction as the product of 
of a complex exponential function and a linear function proportional to the (0,1) sequences. 
These wavefunctions can be seen as standing waves with zero nodes. 
They are an analogous mathematical construction based on physics quantum states 
with some total physics energy.
Such analogous wavefuntion calculations from genome sequences can be applied to 
sonification as in~\cite{Can4} and illustrated next.

\begin{figure}[H]
\begin{center}
\includegraphics[width=12.0 cm]{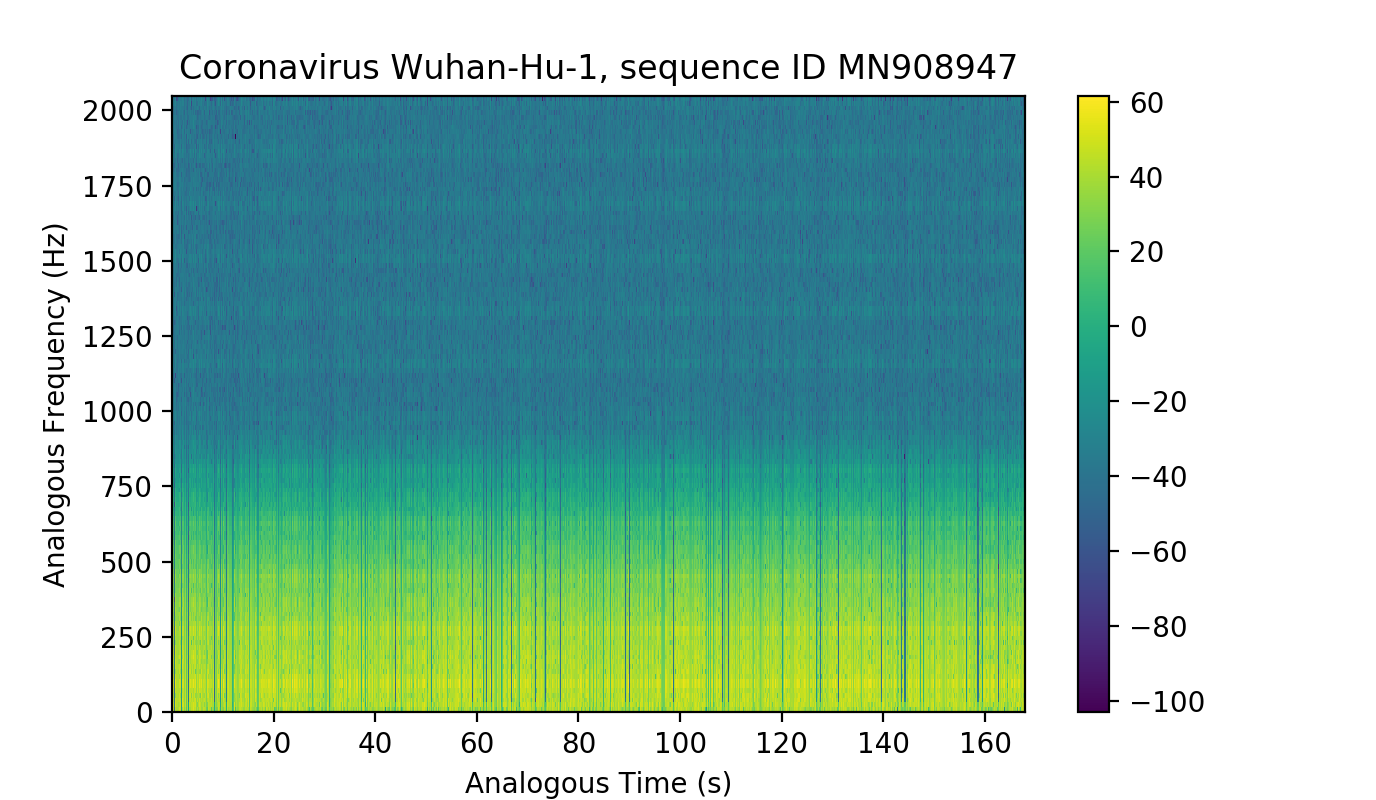}
\caption{
Analogous time-frequency spectrogram of nucleotide bases Adenine
produced from a wav audio file generated via the GenomeBits wavefunction from the first 
Wuhan-Hu-1 coronavirus sequence.
\label{fig5}}
\end{center}
\end{figure}

The analogous time-frequency spectrogram of the nucleotide bases Adenine, showing 
large peak signals over time, is shown in Fig.~\ref{fig5} --generated via separated
GenomeBits wavefunction projections of nucleotide bases A,C,G and T. 
As in~\cite{Can4}, data points in the 
audio curves are fitted to a gaussian in order to obtain a more continuous audio spectrum in 
the waveform reconstruction of the Wuhan-Hu-1 coronavirus sequence (ID MN908947.3). In these 
calculations~\cite{Can4}, we use a similar sample rate: 4096, precision: 16-bit, duration: 2:46.28
min for 681097 sampling: file size: 1.36M: bit rate: 65.5k and sample encoding in one channel: 16-bit. 

A visual inspection of the spectrograms obtained from an audio file could help to identify 
significant virus mutations. Examples of the wav audio files generated via the present extended 
GenomeBits wavefunctions, by transforming the occurrence of nucleotides of the same class along 
the genome sequences can be downloaded from GitHub~\cite{GitHz}. We have artificially shifted these 
curves in frequency by an offset in Hz to obtain a clearer picture when assessing significant 
variations through, e.g., the different densities of colored straight lines. This frequency-shift 
of the analogous audio data simply allows to hear the chirp signal better.

\section{Final remarks and future directions}

This overview summarizes the GenomeBits findings in \cite{Can1,Can2,Can3,Can4} 
outlining the intrinsic signal organization of genomics sequences for different 
coronavirus variants along the pandemic years 2020-2022. All results discussed 
are representative and the plots shown are calculated anew.
The DFT Power Spectrum were calculated for Delta and Omicron variants for 
coronavirus samples from Spain. A difference with respect to Ref.\cite{Can1}   
is that these curves represent results obtained for  
single A,C,G and T nucleotides and not their total sum as previously published.
The results obtained for a 'Order-disorder’ transition are for      
each sequences of A,C,G and T nucleotide of the coronavirus 
Delta and Omicron variants from Spain (a different sequence was used
in Ref.\cite{Can2}). In Ref.\cite{Can3} the Statistical imprints calculations related MPXV disease. 
In this review completely new results are reported for histograms, empirical and theoretical
cumulative distribution curves and the resulting scatter plots for the base nucleotides
A-C versus their complementary base nucleotides T-G for hCoV-19 genome sequences 
of Omicron from USA. The real spectrum of the analogous wavefunction vs. the nucleotide 
base positions corresponds to the 2020 coronavirus Wuhan-Hu-1, China behave like 
sound waves (shown in Figs.4 and 5). The previous Ref.~\cite{Can4} reports results for
representative nucleotide bases from the genome sequence of coronavirus           
Omicron variant. Therein, such curves were artificially shifted in frequency 
by an offset of 400Hz.

In this work, we have considered distinctive patterns of complete coronavirus genome code for single 
strand folded onto itself as available from the GISAID database for the nucleotide bases Adenine, 
Cytosine, Guanine and Thymine. As we discussed in~\cite{Can1}, the coronavirus genome is RNA, not DNA. 
It is possible to correlate symbols used for proteins (polymers of amino acids) to that of nucleic 
acids (polymers of nucleotides). Genomebits numerical results may be relevant to assist in designs 
of the new generation of synthetic messenger ribonucleic acid (mRNA)-based vaccines by a continuous 
surveillance of the evolution of sequence mutations and their capability to replicate. Indirectly, 
such studies could relate to the “central dogma” of modern molecular biology by characterizing the 
processes involved in bringing genetic code from DNA into proteins through mRNA. This is so because 
inter-gene parts of a sequenced genome (revealed potentially by GenomeBits data mine) play most 
likely important roles in the transcription and translation of protein synthesis.

To conclude this comprehensive Bioinformatics overview of the underlying genomics features at the 
nucleotide level from the viewpoint of the physics-based GenomeBits approach let us mention 
potential directions to explore in order to enrich the description of data sequencing.
In theory, one can consider the GenomeBits progression $E_{\alpha,N}(X)$ 
--carrying alternating terms of the independently distributed (0,1) variables $X_{\alpha,k}$,
as the function that contains (most of) the hidden information (or the "secrets") of the genome 
(en-)coding. Processes of binding should be energetically optimized and force driven.
If one accept this interpretation, then by some simple geometrical and physics arguments 
it can be argued that along a 1D continuous positive string, the area under a given coding curve $f(x)$
satisfies
\begin{equation}
\int f(x) dx = \zeta \; \frac{\Delta x}{F}  \;\;\; ,
\end{equation}
where $\zeta$ is energy, $F$ applied force and $\Delta x$ the displacement along our GenomeBit system. 
By considering a spring-like force $F=-K_{_{c}}\cdot\Delta x$, we then approximate for simplicity
\begin{equation}\label{eq:zeta}
\zeta_{_{b}} \approx -K_{_{c}} \int f(x) dx\;\;\; . 
\end{equation}
This relation may correlate a negative binding energy $\zeta_{_{b}}$ with the 
properties contained in the "information" coding experienced through a (discrete "force" series) 
function $f(x) \iff E_{\alpha,k}$ of Eq.(\ref{eq:eq1}), of a supporting genetic sequence (the “system" 
following the information hypothesis of the theoretical framework in Ref.~\cite{Fag22}). 
Instructions to assemble all living organisms must be implicitly included
in the GenomeBits series and the area under these GenomeBits curves can relate to an energy.
In other words the energy-information correlation of Eq.(\ref{eq:zeta}) may give some hints 
to describe the evolution of nucleotides located in the cell nucleus --the fundamental 
unit responsible of live. The genome in a cell carries fundamental information to create 
a living organism, and part of this embedded information is established by meaning and symbols.

All associations discussed in this review are likely to be relevant for the study of genomics
encoding of new sequences in microorganisms. Consequently, the GenomeBits representation 
may be useful to target the evolution of sequences due to natural mutation. One advantage 
for this approach is that sequence data for single A,T,C and G can be handled statistically 
to find their diverse characterizations and determine if the altered coding regions (genes) 
may share similar behavior to those archived and classified during the evolution of a pandemic.
GenomeBits may be useful for the Bioinformatics surveillance behind future infectious diseases 
by means of its simple letter sequence-to-alternating-numerical mapping.

\funding{``This research received no external funding''}

\dataavailability{Binaries for "GenomeBits: A tool for the signal analysis of complete genome sequences",
https://github.com/canessae/GenomeBits/ (Last visited 28/10/2022).}

\conflictsofinterest{``The author declare no conflict of interest.''}

\begin{adjustwidth}{-\extralength}{0cm}

\reftitle{References}

\end{adjustwidth}

\begin{thebibliography}{999}

\bibitem{Vos92}
Voss R.F.,
"Evolution of long-range fractal correlations and 1/f noise in DNA base sequences",
Phys. Rev. Lett. {\bf 68} (1992) 3805.
doi: 10.1103/PhysRevLett.68.3805

\bibitem{Tun16}
Hoang T., Yin Ch., Yau S.S.T.
"Numerical encoding of DNA sequences by chaos game representation with application in similarity comparison",
Genomics {\bf 108} (2016) 134–142.
doi: 10.1016/j.ygeno.2016.08.002

\bibitem{Tho22}
Thornton M., Mcgee M.
"Use of DFT distance metrics for classification of SARS-CoV-2 genomes",
J. Computational Biology {\bf 29} (2022) 453-464.
doi: 10.1089/cmb.2021.0229

\bibitem{Can1}
Canessa E.
"Uncovering signals from the coronavirus genome",
Genes {\bf 12} (2021) 973.
doi: 10.3390/genes12070973

\bibitem{Can2}
Canessa E., Tenze L.
"GenomeBits insight into omicron and delta variants of coronavirus pathogen",
PLoS ONE {\bf 17} (2022) e0271039.
doi: 10.1371/journal.pone.0271039

\bibitem{Can3}
Canessa E.
"GenomeBits Characterization of MPXV",
Genes {\bf 13} (2022) 2223.
doi: 10.3390/genes13122223

\bibitem{Can4}
Canessa E.
"Wave‑like behaviour in (0,1) binary sequences",
Scientific Reports {\bf 12} (2022) 13971.
doi: 10.1038/s41598-022-18360-z

\bibitem{Als22}
Alser M., Lindegger J., Firtina C. et al.
"From molecules to genomic variations: Accelerating genome analysis via intelligent algorithms and architectures",
Computational and Structural Biotechnology J. {\bf 20} (2022) 4579-4599.
doi: 10.1016/j.csbj.2022.08.019

\bibitem{bib-1h}
Canessa E., Tenze L.
Binaries for "GenomeBits: A tool for the signal analysis of complete genome sequences", \\
https://github.com/canessae/GenomeBits/ (Last visited 28/10/2022).

\bibitem{Zha22}
Zhang Y., Wei Y., Yang S. et al.
"Rapid and accurate identiﬁcation of SARS-CoV-2 variants containing E484 mutation", 
The Innovation {\bf 3} (2022) 100183.
doi: 10.1016/j.xinn.2021.100183

\bibitem{Plai21}
Plaisier H., Meagher T.R., Barker D.
"DNA sonification for public engagement in bioinformatics",
BMC Res. Notes {\bf 14} (2021) 273.
doi: 10.1186/s13104-021-05685-7

\bibitem{GitHz}
GitHub: A code hosting platform for version control and collaboration. 
https://github.com/canessae/GenomeBits-Waves (Last visited 28/10/2022).

\bibitem{Fag22}
D’Ariano G.M., Faggin F. 
"Hard problem and free will: An information-theoretical approach", 
in Scardigli F. (eds) Artificial Intelligence Versus Natural Intelligence. Springer (2022). 
doi: 10.1007/978-3-030-85480-5\_5

\end{thebibliography}
\end{document}